\documentclass{jetpl}
\twocolumn

\usepackage{graphicx}
\lat

\title{A comment on ``Different STM Images of the Superstructure on a Clean
Si(133)-6$\times$2 Surface'' (JETP Letters 105 (8), 477-483 (2017))}

\rtitle{A comment on ``Different STM Images\ldots}

\sodtitle{A comment on ``Different STM Images of the Superstructure on a Clean
Si(133)-6$\times$2 Surface'' (JETP Letters 105 (8), 477-483 (2017))}

\author{R.\,Zhachuk$^{+*}$\/\thanks{e-mail: zhachuk@gmail.com}, J.\,Coutinho$^+$}

\rauthor{R.\,Zhachuk R., J.\,Coutinho J.}

\sodauthor{Zhachuk, Coutinho}

\address{$^+$Department of Physics \& I3N, University of Aveiro, Campus Santiago,
3810-193 Aveiro, Portugal\\~\\
$^*$Institute of Semiconductor Physics, pr. Lavrentyeva 13, Novosibirsk
630090, Russia}

\dates{26 June 2017}{*}

\begin{document}

\maketitle

In the recent paper by Teys \cite{tey17}, an atomic model for the
Si(331) reconstructed surface (hereby referred to as T-model) was
proposed on the basis of high-resolution scanning tunneling microscopy
(STM) images. While detailing the virtues against previous and abandoned
models, the author avoids any reference to the rather distinct 8P-model
advocated few weeks earlier by Zhachuk and Teys \cite{zha17}, casting
doubts to his own work. Formulated that way, findings from Ref.~\cite{tey17}
leave readers of JETP Letters with a partial and confusing view of
the problem, and above all, leaves the observations open to ambiguous
interpretation. The 8P-model is also based on STM measurements, and
unlike the T-model, passed through the scrutiny of first-principles
calculations. According to the authors, the 8P reconstruction consistently
described the STM imagery and showed a remarkable low surface formation
energy. 

Since the T-model is solely based on STM data, the above ambiguity
can only be dissipated if we compare 8P and T structures on an equal
foot. This means testing the T-model in terms of surface energy and
STM simulations from first-principles. Using the same procedure as
in Ref.~\cite{zha17}, we found that the T-model is actually unstable.
After atomic relaxation, a Si-Si bond in the surface trimer breaks,
leading to a strong rearrangement of the surface atoms. Not surprisingly,
the resulting simulated constant-current STM image is incompatible
with the experimental analogues shown in Figs.~3(a) and 3(b) of Ref.~\cite{tey17}.
The surface energy of the structure attained after relaxing the T-model
is 8~meV/{\AA} higher than the energy of the 8P structure, much higher
than the typical error bar (below $1$~meV/{\AA}) which allows us to
discriminate surface stability orderings. Combining these figures with
the upper limit for the Si(331) surface energy \cite{zha17}, we conclude 
that according to the T-model, the Si(331) surface should be unstable 
against decomposition into Si(111) and Si(110) facets, in obvious 
contradiction with the observations.

\begin{figure}
\includegraphics[clip,width=8cm]{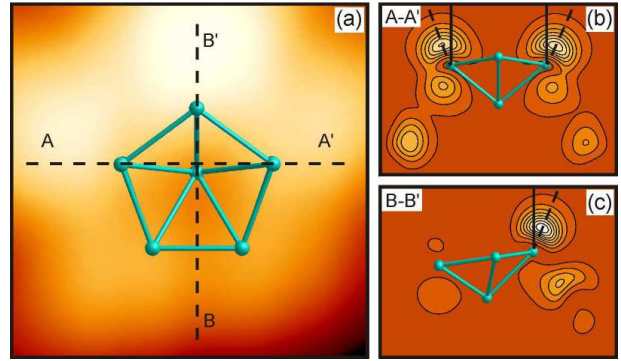}
\caption{\label{fig1}{\bf Figure 1.} (a) Experimental STM image of a pentamer
on Si$(331)$ surface. The atomic model of the pentamer is overlaid.
(b), (c) Contour plots of calculated LDOS isosurfaces on A-A’ and
B-B’ cutting plabes. Experimental and theoretical bias voltages are
$+0.8$~V.}
\end{figure}

In Ref.~\cite{tey17} a critical remark was made against an ancestor
structure of 8P \cite{bat09}, namelly that the pentamer-like features
shown in the experimental STM images are about 1.5-1.8 times larger
than the 5-fold rings of Si atoms from the atomistic model \cite{bat09}.
This brings us to the need of extending the discussion by including
the 8P-model. STM is a technique sensitive to the local density of
\emph{electronic} states (LDOS), rather than the positions of atomic
nuclei \cite{hof03}. It is also obvious that orbitals, particularly
in group-IV semiconductors, can be polarized and maxima of amplitude
are not necessarily centered on atomic nuclei. Figure.~\ref{fig1}(a)
shows the experimental STM image of a pentamer-like feature along
with the Si pentamer model. Dashed lines A-A' and B-B' represent vertical
planes where the LDOS was calculated and the result is shown in Figs.~\ref{fig1}(b)
and \ref{fig1}(c), respectively. The brightest spots in Figs.~\ref{fig1}(b),
(c) indicate a high intensity of the empty LDOS, associated with the
Si radicals at the pentamer vertices. Clearly, the radical states
do not ‘point upwards’, rather making an angle with respect to surface
normal and away from the atomic pentamer. Since the scanning tip hoovers
between 4 and 10~{\AA} above the surface \cite{hof03}, the slanted radicals
project a ‘zoomed’ image of the underlying atomic positions. Combining
the angle as measured from Fig.~\ref{fig1} with the tip height,
we arrive at an estimated zooming factor of about 1.5-2.0 times, thus
clarifying what would be better classified as a magnification effect. 

In conclusion, the present comment reconciles Ref.~\cite{tey17}
with the literature by supplementing the discussion with a missing
and critical account on the stability and electronic structure of
the T- versus 8P-models of Si(331). From first-principles calculations,
we refute both the T-model proposed in Ref.~\cite{tey17}, as well
as the argument used therein against pentamer formation. Conversely, 
we demonstrate that besides showing a very low surface formation energy 
(indeed well below the multi-faceting limit), the 8P-model of the Si(331) 
surface nicely reproduces the STM observations.

This work was funded by the Funda\c{c}ao para a Ci\^{e}ncia e a Tecnologia
(FCT) under the contract UID/CTM/50025/2013, and by FEDER funds through
the COMPETE 2020 Program.


\begin{thebibliography}{99}

\bibitem{tey17}
S.\,A.\,Teys, JETP~Letters {\bf 105}, 477 (2017).

\bibitem{zha17}
R.\,Zhachuk, S.\,Teys, Phys.~Rev.~B {\bf 95}, 041412 (2017).

\bibitem{bat09}
C.\,Battaglia, K.\,Ga\'al-Nagy, C.\,Monney, C.\,Didiot, E.\,F.\,Schwier, M.\,G.\,Garnier, G.\,Onida, P.\,Aebi, Phys.~Rev.~Lett. {\bf 102}, 066102 (2009).

\bibitem{hof03}
W.\,A.\,Hofer, Prog.~Surf.~Sci. {\bf 71}, 147 (2003).

\end{thebibliography}
\end{document}